\documentclass[twocolumn,showpacs,preprintnumbers,amsmath,amssymb,prl]
{revtex4}
\usepackage{graphicx}
\usepackage{dcolumn}
\usepackage{bm}

\begin{document}
\title{Electron density stratification
in two-dimensional structures tuned by electric field}
\author{V.Yu.~Kachorovskii}
\email{kachor.vip1@pop.ioffe.rssi.ru}
\author{I.S.~Lioublinskii}
\author{L.D.~Tsendin}
\affiliation{A.F.~Ioffe Physical-Technical Institute,
26 Polytechnicheskaya str.,
Saint Petersburg, 194021, Russia}
\date{\today}
\begin{abstract}
A new kinetic instability which results in
formation of
charge density waves  is proposed.
A spatial period of arising
space-charge and field configuration is inversely proportional to
electric field and can be tuned by  applied voltage.
The instability  has no interpretation in the  framework of
traditional hydrodynamic
approach, since it arises from modulation of an electron
distribution function  both in coordinate and energy spaces.
The phenomenon can be observed in thin 2D
nanostructures at relatively low electron density.
\end{abstract}
\maketitle
Recent progress in microelectronics is related to great success
in controlled fabrication of low-dimensional semiconductor
systems.  That is why transport properties of semiconductor
nanostructures, both of classical and quantum nature, lately
attract
wide attention.  In this paper we discuss
a new type of purely classical instability, which
 can be observed in 2D nanostructures.
The instability results in a formation  of charge
  density waves (CDW). 
The main feature of the phenomenon is its kinetic nature.
In contrast to usual current and density instabilities in
semiconductors  \cite{konuel,gant,kerner},
the electron
dynamics in the kinetic instability  can not be described
on the basis of the local hydrodynamic
parameters, such as electronic density, drift velocity, and temperature.
Remarkably, the kinetic instability can develop even in the Ohmic regime, 
when the stationary value of electric current is proportional to 
applied voltage.  

Closely related phenomena of the striations formation are widely
known in the gas discharge physics.
The striated discharge has been observed  since M.~Faraday and
is regarded as one of the most typical discharge forms
\cite{raizer,kniga,oleson,nedosp,pekar}.
In spite of this, a consistent
theory of striation is up to now absent.
In the last decades it was realized that for description of
typical striated discharges
the fluid approach fails.
It was shown that the
hydrodynamic description is valid only for very high
electron densities when the collisions between electrons are
frequent enough for the  maxwellization of electron distribution 
function (EDF) \cite{tsenHD,wijacz}.
At  lower
electron densities, occurring in typical gas discharges,
the phenomenon  is of  essentially kinetic
nature. In this case, the EDF perturbation in the striations
is varying both in space, and along the energy axis
\cite{tsendin,muller,ruzh1,rayment,ohe2,golub1},
and it
is impossible to parameterize it in terms of perturbations of
electron density and temperature.  The  kinetic
striations mechanism
was analyzed first
in \cite{tsendin,muller,ruzh1}.
In
\cite{tsendin} it was argued, that
the necessary conditions for  kinetic stratification are: a)
The momentum relaxation is much faster than the energy relaxation; b)
The energy relaxation is mostly controlled by the energy gain in
the external field $F_0$ and strong inelastic collisions with a 
large fixed
energy transfer $W_0$;  c) There should  exist a mechanism
of a weak continuous energy loss.
In a spatially modulated  potential
$U(z)= -F_0 z + \delta U(z),~ \delta
U(z)=\delta U(z+L)$, these 
conditions provide \cite{tsendin,muller,ruzh1}
for the resonant  EDF response  at  $L = L_0/m, $ where $L_0=W_0/F_0,~
m = 1,2,..$.   This "resonant" behavior corresponds to the widely known
empirical Novak's rule \cite{oleson,nedosp,pekar,novak}.
An idea was put forward \cite{tsendin} that under  the
conditions a)-c) the
instability  develops which results in the formation of  CDW
with the same periods $L_0/m$. In  \cite{tsendin,muller,ruzh1},
the electron kinetics
was analyzed only in a given fixed electric potential profile $U(z)$.
However, a complete analysis of instability requires self-consistent
calculation of the
potential  perturbation $\delta U(z,t)$  in terms
of the carrier densities perturbations.  Since the discharge field
depends crucially on the  ion motion and on the  complex
ion generation processes, even a linear instability problem
for the gas discharge plasma is still lacking a self-consistent
solution.

In this paper we will demonstrate that, in principle,
the kinetic  stratification
is also observable in
low-dimensional semiconductor structures \cite{hydro}.
Moreover, it turns out that for the semiconductors a relatively
simple self-consistent analytical solution can be found. The main
simplification follows from the fact, that, in contrast to the
gas discharge, a compensating positive charge
is fixed and homogeneous.
The stratification conditions a), b), and c) can be easily
achieved in
semiconductors.  The momentum relaxation is usually fast
compared to energy relaxation.  The requirements b) and c) are
also usually satisfied, the scattering by optical phonons with
energy $W_0$ and scattering by acoustic phonons working
as strong inelastic and weak quasielastic energy relaxation
mechanisms.
We will show that  effect can be  observed in  2D
quantum wells with small thickness.
  Spatial periods of  arising CDW equal to $L_0/m$  and can 
be tuned by applied voltage.
We   assume that lattice
temperature $T_0,$ as well as Fermi energy,  are
small compared to $W_0$ (in what follows for simplicity
we put  $T_0=0$).
The condition b) requires that
electrons be "hot", and their energies be of the order of
$W_0,$ i.e. the electron gas in this case  is
nondegenerate. We also assume  that electron concentration is
small and neglect electron-electron collisions.

Let us consider the motion of an electron in an external field
$F_0$ assuming for a moment that the only scattering mechanism
is elastic  scattering. This leads to a diffusion in
coordinate space.  As far as energy relaxation processes are
"turned off", the electron is infinitely heated by
the field $F_0$ diffusing over the kinetic energy $W.$
Evidently, this
diffusion is  strictly correlated with the diffusion in the
coordinate space, since the full electron energy $E= W - F_0 z$
is conserved. In fact, an electron diffuses in $(z, W)$ space
along the line $E = const.$ Now, if we take into account
sufficiently intensive optical
phonon emission, the electron motion will be
restricted by a shell $0<W<W_0.$  In the  process of
diffusion with a constant total energy $E,$ the  kinetic energy
increases.  Reaching the point $W=W_0,$ electron loses the energy
$W_0$ and starts a diffusive motion with a lower total energy $E -
W_0.$ A trajectory of the electron in the space $(E,z)$ is shown
on Fig. 1a.
\begin{figure}
\includegraphics{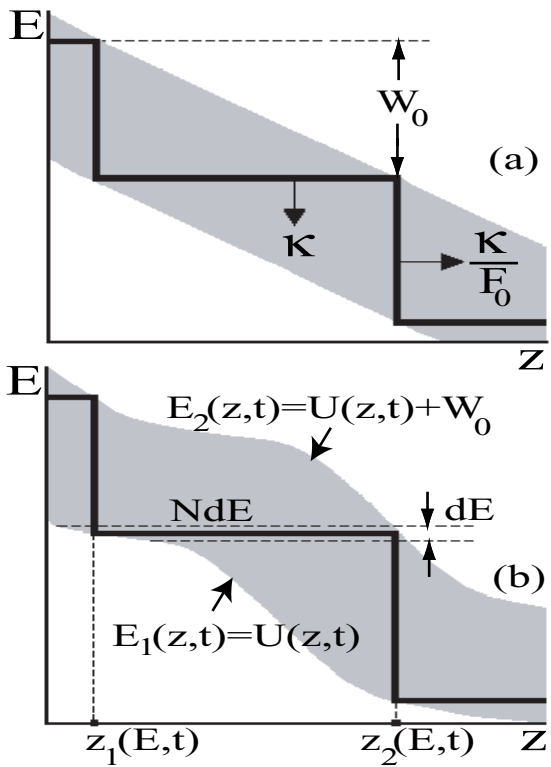}
\caption{\label{Fig1} (a) Motion of an electron in homogeneous field,
$U_0(z)= -F_0 z.$ The  diffusive "staircase"
trajectories slowly drift in $z$ direction with velocity
$\kappa/F_0.$  (b) Motion in the  modulated potential
$U(z,t)= -F_0 z + \delta U(z,t).$}
\end{figure}
The interaction with acoustic phonons plays a role of
a friction force leading to the continuous loss of the electron energy
with the rate $\kappa$ due to spontaneous
emission (since we
have assumed that $T_0 = 0$).  Later we will show that for a 2D
system the rate of  energy loss $\kappa$ does not depend on the
kinetic energy.  There are three different time scales in our
problem: a transport scattering time $\tau,$ a characteristic
time of electron heating by the electric field $\tau_0 \sim L_0^2/D_0$
(here $L_0=W_0/F_0,~D_0 = W_0 \tau/ M$ and $M$ is the  electron 
effective mass),
and a time $W_0/\kappa$ which characterizes a rate  of energy
loss due to the emission of acoustic phonons. We will assume that
\begin{equation}
\tau \ll \tau_0 \ll \frac{W_0}{\kappa} .
\label{ineq}
\end{equation}
Inequality (\ref{ineq})
provides that the acoustic phonon scattering may be considered as a
small perturbation. Due to this scattering the "staircase" diffusive
trajectories move slowly down (along axis $E$) with velocity
$\kappa.$  This means at the same time  that the trajectories
slowly drift with the velocity $s_0=\kappa/F_0$ along axis $z$
(see Fig. 1a).
Since $s_0$ is inversely proportional to the applied field one can say
that the motion of  trajectories demonstrates a negative
differential mobility. It is well known that the negative
differential mobility should lead to the current instability
\cite{konuel,gant}.
However, our case is
more complicated than the usual Gann instability, because one has to follow
the motion of diffusive trajectories instead of the motion
of individual electrons. As for latter ones, their average
drift velocity obeys the usual Ohm's law $v=F_0\tau/M$ and is much
larger than $s_0$ (since the inequalities (\ref{ineq}) may be
rewritten as $ s_0 \ll v \ll \sqrt{W_0/M}$). The fact that
the instability can be observed in the Ohmic regime indicates
that the effect is purely kinetic and can not be described
in terms of hydrodynamic parameters.

As far as the elastic collisions are dominant (see Eq.(\ref{ineq})),
the EDF is almost isotropic \cite{denis},
$f(z,W,\varphi,t) \approx f_i(z,W,t) +
f_a(z,W,t)
\cos(\varphi).$
Here $f_i$
is an isotropic part of EDF, $ f_a \cos(\varphi)$
is a small anisotropic correction, and $\varphi $
is an angle between electron
velocity and applied field. Denote $J(z,W,t)= \sqrt{W/4M}f_a.$
The equations for $f_i$ and $J$  can
be written as follows \cite{konuel,gant,denis}
\begin{equation}
J=-D(W)\left(\frac{\partial f_i}{\partial z} +F \frac{\partial f_i}
{\partial W}\right),
\label{J0}
\end{equation}
\begin{equation}
\frac{\partial f_i}{\partial t}+ \frac{\partial J}{\partial z}
+F \frac{\partial J}{\partial W}=
\kappa \frac{\partial f_i}{\partial W},
\label{f0}
\end{equation}
where $D(W)=W\tau/M$ is an energy dependent diffusion
coefficient. For simplicity we assume that $\tau$ is energy independent
and, consequently, $D(W)$ is proportional to $W.$
The boundary conditions for Eqs. (\ref{J0}), (\ref{f0}) read
\begin{equation}
f_i\vert_{W=W_0} =0, \qquad (FJ-\kappa f_i)\vert_{W=0}=FJ\vert_{W=W_0}.
\label{b1}
\end{equation}
Here $F(z,t)= - \partial U(z,t)/\partial z, $ and $U(z,t)$ is
a potential energy, which includes both the self-consistent potential
created by electrons and the external potential $U_0(z)=
-F_0z.$   Condition $ f_i\vert_{W=W_0} =0$ corresponds to
the limit of a very strong interaction with optical phonons
("black wall" condition). The second boundary condition
is related to the conservation of the number of particles
in inelastic collisions \cite{standard}.
Eqs. (\ref{J0}), (\ref{f0}), and (\ref{b1}) at $\kappa=0$
have a homogeneous stationary solution
\begin{equation}
J=J_0 =\frac{ n_0 v}{W_0},~~~
f_i=f_{i0}= \frac{n_0}{W_0}\ln \left(\frac{W_0}{W}\right),
\label{stats}
\end{equation}
where $n_0$ is the stationary electron concentration (we assume the
following normalization
$\int^{W_0}_0 f_{i0} dW =n_0 $).  Since $\int^{W_0}_0 J_0 dW =n_0 v$,
solution (\ref{stats}) corresponds to the 
Ohmic regime.  
According to
\cite{tsendin} we
rewrite Eqs. (\ref{J0}), (\ref{f0}) in variables $(E, z,t),$
where $E = W + U(z,t)$ is a full energy of a particle.  The
result is given by
\begin{equation}
J=-D(E- U)\frac{\partial f_i}{\partial z},
\label{J1}
\end{equation}
\begin{equation}
\frac{\partial f_i}{\partial t}+ \frac{\partial J}{\partial z}
=
(\kappa -\frac{\partial U}{\partial t}) \frac{\partial f_i}{\partial E}.
\label{f1}
\end{equation}
The motion of a particle in the space of new variables is
restricted by the curves
$E=E_1(z,t)=U(z,t), ~E=E_2(z,t)=W_0+ U(z,t)$ (See Fig. 1b).
The
boundary conditions (\ref{b1}) can be rewritten as
\begin{eqnarray}
f_i\vert_{E=E_2(z,t)}=0,
\nonumber \\
  (FJ-\kappa f_i)\vert_{E=E_1(z,t)}=
FJ\vert_{E=E_2(z,t)}.
\label{eb1}
\end{eqnarray}
Since the total energy of an electron changes slowly
(with the characteristic
time $W_0/\kappa$), it will be useful to introduce an electron
density distribution over the  $E$ axis 
\begin{equation}
N(E,t) = \int_{z_1(E,t)}^{z_2(E,t)} dz f_i,
\label{N}
\end{equation}
where $z_1(E,t),~z_2(E,t)$ are the inverse functions of
$E_1(z,t),~E_2(z,t),$ correspondingly,
and the value $NdE$ represents
the number of electrons on  "staircase" trajectories
restricted by  $E$ and $E+dE$ (see Fig. 1b).
The stationary value of $N$ is given by $N_0 = n_0/F_0.$
Introducing the notation $I_{E}(t)= J(z_2,E, t)$ (the stationary
 of $I_E$ being equal to $J_0$ ) we find from
Eq. (\ref{f1})
\begin{equation}
J(z_1, E,t)= I_E(t) + \int_{z_1}^{z_2} dz \left(
\frac{\partial f_i}{\partial t} -
(\kappa -\frac{\partial U}{\partial t}) \frac{\partial f_i}{\partial E}
\right).
\label{Jz1}
\end{equation}
Taking into account that  $\partial
z_1/ \partial E = - 1/ F(z_1,t)$, \newline $\partial z_1/ \partial t =
(\partial U/ \partial t)/ F(z_1,t)$
and using  Eqs. (\ref{eb1}), (\ref{Jz1}) we obtain the continuity-like
equation that governs the electron motion over the axis of total
energy
\begin{equation}
\frac{\partial N}{\partial t} - \frac{\partial }{\partial E}\left(N
\left[\kappa - \left< \frac{\partial U}{\partial t}\right>\right]\right)
=I_{E +W_0}(t) - I_{E}(t).
\label{cont}
\end{equation}
Here the angle brackets mean  averaging  over $z$
\begin{equation}
\left< \frac{\partial U}{\partial t}\right> = \frac{1}{N}
\int_{z_1}^{z_2} dz\frac{\partial U}{\partial t} f_i(z,E)
\label{<>}.
\end{equation}
%

Next we consider  the  deviations from the
stationary solution in the linear approximation.
A small periodic over coordinate modulation  of  the  potential
$U-U_0=\delta U_q
\exp(-i\omega t + i qz )$
induces the energy dependence of
quantities $I_E,~ N$ in forms
$I_E-J_0=\delta I_q\exp(-i\omega t - i qE/F_0 )$,
$N-N_0=\delta N_q \exp(-i\omega t  - i qE/F_0)$.
We will demonstrate that for $q \approx q_m = \pm 2\pi m/L_0$ 
(where  $m=1,2,..$)  the imaginary part of $\omega$ is positive which implies
that a stationary solution (\ref{stats}) is unstable.
For $q=q_m$  the solution is periodic
function of energy with a period $W_0/m$ and
$I_{E+W_0}(t)=I_{E}(t)$. Then linearization of
Eq. (\ref{cont}) yields
\begin{equation}
\omega_m =
\frac{\kappa}{F_0 + \Delta F_m} q_m,
\label{w(q)}
\end{equation}
where 
$\Delta F_m =
- iq_m N_ 0 \left< \delta U_m\right>/
\delta N_m.$
We see that the physics of the problem is governed
by the only parameter $\left< \delta U_m\right>/
\delta N_m$ (the subscript $m$ implies that all quantities are taken
at $q=q_m$). This parameter has a transparent physical meaning
of a response of
the averaged 
potential with respect to a small variation
of electron density in energy space $\delta N_m.$
The instability (${\rm Im} (\omega_m) >0 $) occurs,
when $ Re \left< \delta U_m\right>/
\delta N_m >0$.
In order to find this parameter one should go beyond
the averaged kinetic equation (\ref{cont}) and solve
Eqs. (\ref{J1}), (\ref{f1}) together with  Poisson equation.
As long as Eq. (\ref{w(q)}) is already proportional to a small
parameter
$\kappa,$ one can simplify the  solution of
Eqs. (\ref{J1}),  (\ref{f1})
assuming that $\kappa =0$, and neglecting $\partial f_i/
\partial t$ and $\partial U/
\partial t$ (since Eq. (\ref{w(q)})
provides that $\omega \sim \kappa$).
Then Eq. (\ref{f1})  reduces to $\partial J/ \partial z = 0,$
which implies that $J(z,E,t)=I_E(t).$ As a result, Eq. (\ref{J1}) yields
\begin{equation}
f_i(z,E,t)= I_E(t) \int_z^{z_2}\frac{dz^{\prime}}{D(E-U(z^{\prime},t))}.
\label{fJ2}
\end{equation}
The small variation of the distribution function $\delta f_i$
can be found by linearization of
this equation
with respect to $\delta I_m, \delta U_m, $
the functions $z_1$ and $z_2$ being also linearized.
The Poisson equation
gives us
a proportionality between $\delta U_m$ and  the
Fourier transform $\delta n_m$ of the variation of
electron concentration
\begin{equation}
\delta n(z,t) = \delta \int^{E_2}_{E_1} dE  f_i.
\label{dn}
\end{equation}
Here the variation
includes the
variation of $\delta f_i$ as well as variation of the integration limits
$E_1(z,t)$ and $E_2(z,t).$
In this paper we restrict ourselves to the case of 2D
semiconductor quantum well, assuming that the
dielectric constant $\epsilon$ is the same both
inside and outside  the  quantum well.
For such structure
\begin{equation}
\delta U_m =
\frac{ 2 \pi e^2}{ \epsilon \vert q_m \vert}\delta n_m .
\label{C}
\end{equation}
Using  Eqs. (\ref{dn}), (\ref{C})  and  linearized Eqs. (\ref{N}), 
(\ref{fJ2}) one can find the relation
between $\delta N_m$ and $\delta U_m.$
To calculate the parameter $\Delta F_m,$
one should also average the variation of potential
$\delta U_m \exp(iq_m z)$ with the stationary distribution
function $f_{i0}(E,z)$
(since we solve the problem in a linear approximation).
After cumbersome but rather straightforward calculations finally
we get
\begin{equation}
\omega_m = s_0 q_m
+ i\frac{\kappa}{W_0}\frac{\lambda_m
\vert \alpha_m \vert^2 }{1 + \lambda_m \alpha_m},
\label{wm}
\end{equation}
where
$\alpha_m=\int_0^{L_0} dy(1-\exp(iq_m y))/y,~
 \lambda_m=
e^2 n_0/m\epsilon F_0.$
It is easy to check that for any $m$, ${\rm Im}(\omega_m)>0.$ Thus, 
for $q=q_m$
a stationary solution is unstable. For a weak field, $\lambda_m
\gtrsim 1,$ the increment is field independent,
${\rm Im} (\omega_m) \sim \kappa/W_0.$
One can show that for $q\approx q_m$
the spectrum reads
\begin{equation}
\omega(q)=\omega_m + (q-q_m)v - i\frac{D^*}{4}(q-q_m)^2,
\label{wq}
\end{equation}
where
$\displaystyle{ D^* = D_0
\left(1+ \frac{\lambda_m \alpha_m^*}{1+\lambda_m \alpha_m}
\frac{2}{iq_m L_0}     \right)}$ (we neglected small corrections of the order of $\kappa$
to $v$ and $D^*$).
This implies that instability exists only in a small vicinity of
$q_m$ (see Fig. 2).
\begin{figure}
\includegraphics{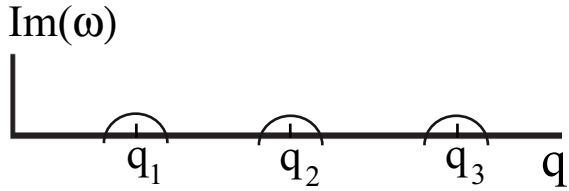}
\caption{\label{fig2} Instability increment as a function of $q$.
Instability regions  correspond to $q \approx q_m= \pm 2\pi m/L_0$}.
\end{figure}
This instability should lead to the formation of
CDW with the periods $L_0/m.$

Next we  discuss a possibility of  observation of the
effect.
The instability increment
is proportional to     the rate of  energy loss $\kappa,$ which
can be calculated for electrons in 2D quantum well in full
analogy with the 3D case \cite{gant}.
For the case of infinitely deep rectangular quantum well of 
 width $a$, calculations  yield
\begin{equation}
\kappa = \frac{C_0^2 \pi^2 M}{\rho a^3 \hbar}.
\label{kappa}
\end{equation}
Here $C_0$ is a deformation potential constant, $\rho$ is  density
of the crystal.
This result justifies our assumption that $\kappa$
does not depend on
electron kinetic energy.
Also we see that  $\kappa$  rapidly increases with decreasing  $a$.
The law $\kappa \sim a^{-3}$ can be
understood from simple  estimates.
The  momentum transfer from electron to
phonon in the direction perpendicular to the quantum well
is of the order of $ \hbar / a.$ Emission of such a
phonon  leads to the energy loss  $\sim \hbar S/a,$ where $S$ is the
sound velocity.   The energy loss rate by the emission of
longitudinal phonons may be neglected  due to a small
factor $k_{||} a,$
where $k_{||}$ is the in-plane wave vector of 2D electron.
We find that $\kappa$ is proportional
to the integral over $dq_z$ of the product of energy loss $\hbar S/a$ by the
squared matrix element $V_q^2 \sim q \sim 1/a.$ The upper limit of the
integral
is  of the same order, of $1/a,$ yielding $\kappa \sim a^{-3}.$
This implies that the instability
is more likely to be observed in thin 2D
structures.
On the other hand, the instability is suppressed by
the electron-electron collisions, which lead to maxwellization
of the EDF.
Thus, the instability  condition is given by
${\rm Im}(\omega_m) >
1/\tau_{ee},$  where $\tau_{ee}$ is the characteristic time of
the electron-electron scattering.
  Crude estimate  of $\tau_{ee}$ for  hot
electrons with characteristic energy
$W_0$ gives  $\tau_{ee}^{-1} \sim e^4 n_0/\epsilon^2 \hbar  W_0.$
Having in mind Eqs. (\ref{ineq}), (\ref{wm}), one can see that for
low electron densities $e^4 n_0/\epsilon^2 \hbar  < \kappa, $
a certain field interval exists, in
which the instability can be observed. Simple estimates for
GaAs and GaN show that for thin quantum wells, 
  $a \approx 30
A^{\rm o}, $ the electron concentration is restricted by
small but quite reasonable value  $\sim 10^{10} sm^{-2}.$

In  conclusion, we have presented 
a self-consistent theory of
kinetic stratification. We have shown that the spatial periods of 
 strata equal to
 $W_0/F_0 m$ and can be tuned by applied voltage.

The work was supported by  RFBR  and  INTAS.


\begin{thebibliography}{20}
\bibitem{hydro} The  hydrodynamic strata in semiconductors were
discussed in \cite{kerner}.
\bibitem{standard} This condition ensures that
integration of Eq.(\ref{f0})  over $dW$
yields
a standard continuity equation (see \cite{tsendin}).
\end{thebibliography}
\end{document}